\documentclass[eqsecnum,superscriptaddress,showpacs]{revtex4}
\usepackage{bm}
\usepackage{graphics}
\usepackage{latexsym}
\begin{document}
\title{Nuclear Potential with Two Pion Exchange}

\author{Takehisa Fujita}\email{fffujita@phys.cst.nihon-u.ac.jp}
\author{Naohiro Kanda}\email{nkanda@phys.cst.nihon-u.ac.jp}
\affiliation{Department of Physics, Faculty of Science and Technology, 
Nihon University, Tokyo, Japan}
\author{Sachiko Oshima\footnote{On leave of absence from : Department of Physics, 
Faculty of Science, University of Tokyo, Japan}
}\email{oshima@nt.phys.s.u-tokyo.ac.jp}
\affiliation{Fakult\"at f\"ur Physik, Universit\"at Duisburg-Essen, Germany}

\date{\today}%

\begin{abstract}

We carefully calculate the nucleon-nucleon interaction due to two pion 
exchange processes by properly evaluating the corresponding Feynman diagrams. 
In the estimation, we have made no approximation, and instead, we carry out 
the numerical integrations of three Feynman parameters. It is found that the two pion 
exchange potential gives rise to the attractive force which corresponds to 
the effective scalar meson with its mass of $m_s\simeq 4.7 m_\pi$ and its strength of 
${g_s^2\over 4\pi}\simeq 1.45 $. There is a strong isospin dependence of 
$(\bm{\tau}_1\cdot \bm{\tau}_2)^2 $ which cannot be simulated by the one boson 
exchange model calculations.

\end{abstract}

\pacs{21.30.-x,13.75.Cs }

\maketitle

\section{Introduction}
The structure of the nucleus can be described once the nucleon-nucleon interactions 
are properly known. Indeed there are already sufficiently large number of works 
available for the determination of the  nucleon-nucleon potential 
\cite{mach1,mach,paris,paris1}. The most popular nuclear interaction may be obtained 
by one boson exchange potential (OBEP) \cite{obep68,sakurai,bohr} where exchanged bosons 
are taken from experimental observations. In this case, the masses and the coupling 
constants of the exchanged bosons are determined from various methods, partly 
experimentally and partly theoretically. 
The discussions of the determination of these parameters may have some ambiguities, 
but one can see that the basic part of the nuclear force can be well understood 
until now.  

However, there is one important problem which is not solved yet completely. This is 
related to the medium range attraction of the nucleon-nucleon potential, and this is 
normally simulated by the effective scalar meson exchange process. It is clear that 
there is no massive scalar meson in nature and, therefore, the artificial introduction 
of the scalar meson is indeed a theoretical defect of the one boson exchange model. 
This is indeed a homework problem for many years of nuclear physics research. 
However, this important problem is left unsolved for a long time since 
many of the nuclear theorists moved and got interested in the quark model calculations 
of the nucleon-nucleon interaction. By now, it becomes clear that the evaluation 
of the QCD based model has an intrinsic difficulty due to the gauge dependence 
of the quark color charge \cite{fujita} and this strongly suggests that 
the meson exchange approach is indeed a right direction  of the nuclear force 
calculations. 

The origin of the medium range attraction has been discussed extensively, but until now 
there is no clear answer to this problem. Mostly, people believe that the medium range 
attraction may well be simulated by the second order calculation of one pion 
exchange potential in which the intermediate $\Delta-$resonance state is included 
\cite{hol,val}. However, it is also known that this cannot give rise to the sufficiently 
large contributions to the medium range attractions. Also, there are many 
calculations of the nucleon-nucleon interaction due to the two pion exchange 
processes \cite{wort,hara,part}, and this may indeed give rise to the medium range 
attraction even though until now there is no clear cut evaluation which can 
isolate the nuclear force contribution to the medium range attraction. 

In this paper, we present a careful calculation of the two pion exchange processes,  
and in the evaluation, we have made no approximation. We first evaluate 
the corresponding Feynman diagrams which contain the integration of the four momentum 
$k$. This integration can be carried out exactly by introducing the Feynman parameters, 
and thus we are left with the integrations of the three Feynman parameters. This can be 
only carried out numerically, and we find that the corresponding T-matrix 
can be expressed reasonably well, at least, up to $\bm{q}^2 \simeq (6m_\pi)^2$ as 
$$ T(\bm{q}) \simeq -g_s^2(\bm{\tau}_1\cdot \bm{\tau}_2)^2 
{1 \over \bm{q}^2+m_s^2 } \eqno{(1.1)}  $$
where
$$ m_s \simeq 4.7 m_\pi, \ \ \ {g_s^2\over 4\pi} \simeq  1.45 
\ \ \ \ \ {\rm with} \ \ \ \ \ \  {g_\pi^2\over 4\pi}\simeq 8  \eqno{(1.2)}  $$
where the mass $m_s$ and the coupling constant $g_s$ should have a small increase 
( up to $\sim$ 20 \% ) as the function of $\bm{q}^2$. 
Here, $g_\pi$ and $m_\pi$ denote the $\pi NN$ coupling constant and pion mass, 
respectively. Therefore, the predicted values of the mass and the coupling constant 
of the effective $\sigma$ meson from the two pion exchange diagrams are given 
for the $T=0$ channel as
$$m_\sigma=m_s \simeq 650  \ {\rm MeV}, \ \ \ \ \  {g_\sigma^2\over 4\pi}
=9\times {g_s^2\over 4\pi} \simeq 13, \ \ \ \ \ 
 (T=0, \ \ {\rm NN- state})  \eqno{(1.3)}  $$ 
which should be compared with the phenomenological values of the $\sigma$ meson mass and 
coupling constant as determined by fitting to the nucleon-nucleon scattering 
data for the $T=0$ channel \cite{mach1,mach}
$$ m_\sigma \simeq 615 \ {\rm MeV}, \ \ \ {g_\sigma^2\over 4\pi} \simeq  11.7 . 
 \eqno{(1.4)}  $$
They indeed agree with each other at the quantitative level. 
The physical reason why the two pion exchange process becomes quite large 
can be easily understood. The one pion exchange potential is suppressed 
due to the pseudo-scalar interaction with nucleon, and thus the one pion exchange 
process in the second order diagram is relatively weak. Indeed, one knows that 
the one pion exchange calculation should pick up the small component in the nucleon 
Dirac wave function. However, the two pion exchange diagram has no such suppression 
and thus it can give rise to the largest contribution to the nucleon-nucleon potential. 
Since it is the fourth order process, it turns out to be an effective scalar 
interaction which is always attractive. In addition, it should be important to note 
that there is no double counting problem in this calculation since the two pion 
exchange process does not contain any one pion ladder type diagrams. 

\section{Two Pion Exchange Process}
In addition to the one boson exchange processes, one should consider 
the two pion exchange diagrams in order to obtain a proper nucleon-nucleon 
interaction. There are of course some calculations of the two pion exchange 
diagrams \cite{gross}, but until now there is no solid calculation of 
the two pion exchange potential which is compared to the observed data.  
However, one can easily convince oneself that the fourth order process 
involving the four $\gamma_5$ interactions is not suppressed at all, in contrast 
to the one pion exchange diagram where the $\gamma_5$ coupling is indeed 
suppressed by the factor of $ {m_\pi\over M}$ with $M$ denoting the nucleon mass, 
which is basically due to the parity mismatch. 
Therefore, it should be very important to calculate the two pion exchange 
process properly in order to understand the medium range attraction of 
the nucleon-nucleon interaction. 

Now, the evaluation of the two pion exchange Feynman diagram is done 
in a straight forward way \cite{gross,bd}, and we find the corresponding T-matrix as
$$ T=ig_\pi^4 (\bm{\tau}_1\cdot \bm{\tau}_2)^2 \int {d^4k \over (2\pi)^4} 
i\gamma_5^{(1)} {1\over k^2-m_\pi^2+i\varepsilon} {1\over (p_1-k)^\mu \gamma_\mu^{(1)}-M 
+i\varepsilon } i\gamma_5^{(1)}   $$
$$ \times i\gamma_5^{(2)} {1\over (q-k)^2-m_\pi^2+i\varepsilon} 
{1\over (p_2+k)^\mu \gamma_\mu^{(2)}-M +i\varepsilon } i\gamma_5^{(2)}  \eqno{(2.1)}   $$
where $p_1$ (${p'}_1$) and $p_2$ (${p'}_2$)  denote the initial (final) four momenta 
of the two nucleons, and $q$ is the four momentum transfer which is defined as 
$q=p_1-{p'}_1$. Here, we have ignored the crossed diagram which is much smaller 
than eq.(2.1). By noting 
$$ (\gamma_5^{(1)})^2=1, \ \  (\gamma_5^{(2)})^2=1, \ \ 
\gamma_5 \gamma^\mu =-\gamma^\mu \gamma_5 $$
we can rewrite eq.(2.1) as
$$ T=ig_\pi^4(\bm{\tau}_1\cdot \bm{\tau}_2)^2 \int {d^4k \over (2\pi)^4} 
{1\over k^2-m_\pi^2} {1\over (q-k)^2-m_\pi^2} 
 \times {(p_1-k)^\mu \gamma_\mu^{(1)}-M \over (p_1-k)^2-M^2 } \times 
{(p_2+k)^\mu \gamma_\mu^{(2)}-M \over (p_2+k)^2-M^2} . \eqno{(2.2)}   $$
Now, we introduce the integration trick in terms of Feynman parameters $x,y,z$ as
$${1\over abcd}=6\int_0^1 dx \int_0^x dy \int_0^y dz 
{1\over \left[ a+(b-a)x+(c-b)y+(d-c)z \right]^4 } . $$ 
Further, we assume that the nucleons at the initial state are on the mass shell 
$$ (p \llap/_1-M)u(p_1) = 0, \ \ \ \  (p \llap/_2-M)u(p_2) = 0  $$
and therefore we also find
$$ \bar{u}({p'}_1) q^\mu \gamma_\mu u(p_1) =0 . $$
 In addition, we take the non-relativistic limit for the nucleon motion and thus 
obtain
$$ T \simeq -6ig_\pi^4(\bm{\tau}_1\cdot \bm{\tau}_2)^2
 \int_0^1 dx \int_0^x dy \int_0^y dz \int {d^4k \over (2\pi)^4} 
{{1\over 4}k^2+M^2 (2z-y)^2 \over (k^2-s)^4 } 
 \eqno{(2.3)}   $$
where $s$ is defined as
$$ s=q^2\left( (y-x)^2+y-x\right)+M^2 (2z-y)^2+m_\pi^2 (1-y) . \eqno{(2.4)}  $$
The momentum integration of $k$  can be easily done, and we find
$$ T \simeq -{g_\pi^4\over 32\pi^2}(\bm{\tau}_1\cdot \bm{\tau}_2)^2 
\int_0^1 dx \int_0^x dy \int_0^y dz 
\left[ {1\over s} -{2M^2 (2z-y)^2\over s^2}
\right]  . \eqno{(2.5)}  $$
This three dimensional integration of $x,y,z$ can be done only numerically, and 
the calculated result can be well fit by the following shape 
$$ T \simeq -(\bm{\tau}_1\cdot \bm{\tau}_2)^2 {g_\pi^4\over 32\pi^2}\times   
{A\over \bm{q}^2+m_s^2 } \eqno{(2.6)}  $$
where $A$ and $m_s$ are found to be
$$ A \simeq 0.57, \ \ \ \ m_s \simeq 4.7 m_\pi \simeq 650 \ {\rm MeV}. $$
Here, we replace the four momentum transfer of $q^2$ as
$$ q^2 =q_0^2-\bm{q}^2 \simeq -\bm{q}^2 $$
since we may use the static approximation to a good accuracy
$$ (q_0)^2=\left( \sqrt{M^2+\bm{p}_1^2 }-\sqrt{M^2+\bm{{p'}}_1^2 }
\right)^2 \simeq {1\over 4M^2}(\bm{p}_1^2-\bm{{p'}}_1^2)^2 << \bm{q}^2 .  \eqno{(2.7)}   $$
If we take the value of the $\pi NN$ coupling constant as 
${g_\pi^2\over 4 \pi} \simeq 8 $, then we find
$$ T \simeq -(\bm{\tau}_1\cdot \bm{\tau}_2)^2{g_s^2 \over \bm{q}^2+m_s^2 } \eqno{(2.8)}  $$
where
$$   {g_s^2\over 4\pi} \simeq {1\over 4\pi}\times 
{g_\pi^4\over 32\pi^2} \times 0.57 \simeq 1.45  \eqno{(2.9)}  $$
which are consistent with the values determined from the nucleon-nucleon 
scattering experiments. 

It should be important to note that the present calculation suggests that 
the $T=0$ channel of the nucleon-nucleon interaction is very strong 
in comparison with the  $T=1$ case. This means that the proton-neutron interaction 
is much stronger than one would naively expect. 
This fact is, of course, quite well known to nuclear structure physicists since they know 
from the fitting of the spectrum to experiments that the proton-neutron force in 
the residual nuclear interaction is quite strong while the interactions between 
identical nucleons are rather weak.

\section{Conclusions}

We have presented a new calculation of the old type Feynman diagram in the two pion 
exchange processes. The result is quite interesting since the corresponding T-matrix 
of the two pion exchange diagrams turns out to be just similar to the T-matrix of 
the effective $\sigma$ meson exchange case where we obtain the corresponding effective 
$\sigma$ meson mass $m_\sigma \simeq 650$ MeV and the effective coupling constant 
${g_\sigma^2\over 4\pi}\simeq 13 $ \ for $T=0$ channel. These values of the mass and 
the coupling constant are found to agree with those values ($m_\sigma \simeq 615$ MeV, 
\   ${g_\sigma^2\over 4\pi}\simeq 11.7 $) which are determined from the OBEP analysis 
of the nucleon-nucleon scattering experiments. 

The result of the present study may invoke some further investigations 
of the nucleon-nucleon potential. In this calculation, we have employed 
the value of the $\pi NN$ coupling constant $g_\pi$ which is determined 
from the pion decay process into two photons, $\pi^0 \rightarrow \gamma + \gamma$,  
since its value is known to be ${g_\pi^2\over 4\pi}\simeq 8$ 
\cite{nishi,kaft}. 
On the other hand, people use the $\pi NN$ coupling constant $g_\pi$ 
in nucleon-nucleon potential which is somewhat larger than the above value, 
almost by a factor of two. 
This should be closely related to the mass and the coupling constant of 
the effective $\sigma$ meson. Now, the property of the effective  $\sigma$ 
meson is determined by the two pion exchange process and thus we should 
reconsider the one boson exchange potential from the new point of view.  
It should be quite interesting to know to what extent we can understand 
the nucleon-nucleon scattering data with the new constraints of the 
$\pi NN$ coupling constant. However, it is, at the same time, clear that the two pion 
exchange process cannot be simulated by the effective  $\sigma$  meson exchange 
as far as the isospin dependence is concerned.

\end{document}